\documentclass[aip,pof,reprint,floatfix,amsfonts,amssymb,amsmath,superscriptaddress]{revtex4-1}
\usepackage{graphicx}
\usepackage{wasysym}
\usepackage{amssymb}
\begin{document}
\title{Design and Experimental Validation of a Ducted Counter-rotating Axial-flow Fans System}

\author{H. Nouri}
\affiliation{Arts et Metiers ParisTech, DynFluid, 151 boulevard de l'H{\^o}pital, 75013 Paris, France.}
\author{F.~Ravelet}
\email{florent.ravelet@ensta.org}
\affiliation{Arts et Metiers ParisTech, DynFluid, 151 boulevard de l'H{\^o}pital, 75013 Paris, France.}
\author{F. Bakir}
\affiliation{Arts et Metiers ParisTech, DynFluid, 151 boulevard de l'H{\^o}pital, 75013 Paris, France.}
\author{C. Sarraf}
\affiliation{Arts et Metiers ParisTech, DynFluid, 151 boulevard de l'H{\^o}pital, 75013 Paris, France.}
\author{R. Rey}
\affiliation{Arts et Metiers ParisTech, DynFluid, 151 boulevard de l'H{\^o}pital, 75013 Paris, France.}

\begin{abstract}
{\it An experimental study on the design of counter-rotating axial-flow fans was carried out. The fans were designed using an inverse method. In particular, the system is designed to have a pure axial discharge flow. The counter-rotating fans operate in a ducted-flow configuration and the overall performances are measured in a normalized test bench. The rotation rate of each fan is independently controlled. The relative axial spacing between fans can vary from $17\%$ to $310\%$. The results show that the efficiency is strongly increased compared to a conventional rotor or to a rotor-stator stage. The effects of varying the rotation rates ratio on the overall performances are studied and show that the system has a very flexible use, with a large patch of high efficient operating points in the parameter space. The increase of axial spacing causes only a small decrease of the efficiency}
\end{abstract}
\maketitle        
\section{\label{sec:intro}Introduction}
Early studied in the 1930's \cite{Lesley1933}, the counter-rotating machines arouse a greater interest in the turbomachinery field, particularly for their potential improvement of the efficiency with respect to conventional machines by recovering kinetic energy from the front rotor exit-flow and by adding energy to the flow. The first counter-rotating machines have appeared in aeronautic \cite{Lesley1933,Playle} and marine applications \cite{Gunsteren,Chen1989} in open configuration. 

Conventional designs of high speed counter-rotating fans are based on quite expensive methods and require a systematic coming and going between theoretical methods -- such as the lifting line theory or the strip-analysis approach \cite{Playle}-- and CFD analysis \cite{Bechet}. Moreover, the axial spacing, which has a major role on the rotors interaction and consequently on the noise \cite{Holste1997,Blandeau}, is a key parameter to find a compromise between high aerodynamic and good acoustic performance for high speed fans \cite{Bechet}. In order to reduce this interaction, the axial spacing of high speed fans has to be relatively large, resulting in a decrease in the aerodynamic performance \cite{Bechet}. For the same reason, the rear rotor (RR) diameter has to be smaller (about 10$\%$ according to \cite{Bechet,Peters2012}) than the front rotor (FR) diameter to reduce interaction between the FR tip vortex and the RR blade tip.

Contrary to that, in the case of low speed fans axial spacing could be shortened using the benefit of a relatively low rotor interaction. Therefore these machines see a revival of interest in several distinct configurations --open and ducted flows, shrouded or not shrouded rotors-- in various subsonic regime applications \cite{Shigemitsu2009,Shigemitsu2010,Pin2011}. 

Recent research work dealt with the effects of global parameters like rotation speed ratio \cite{Shigemitsu2005}, local phenomena such as tip vortex flows \cite{Xu} and improvement of cavitation performance for pumps \cite{Shigemitsu2009}. All previous studies have shown the benefit of RR in improving the global efficiency and in increasing the operating flow-rate range while maintaining high efficiency. The counter-rotating systems (CRS) moreover allow to reduce the fans diameter and/or to reduce the rotation rate. More axial spacing is needed compared to one simple fan, but not much more than a rotor-stator stage. However, it requires a more complex shaft system. Another interesting feature of CRS is that it makes it possible to design axial-flow fans with very low angular specific speed $\Omega=\frac{\omega\sqrt{Q}}{(\Delta p_t/\rho)^{3/4}}$ with $\omega=\frac{\omega_{rotor1}+\omega_{rotor2}}{2}$ the mean angular velocity, $Q$ the flow rate, $\Delta p_t$ the total pressure rise, and $\rho$ the fluid density. With such advantages, the CRS becomes a very interesting solution and the interaction between the rotors needs to be better understood in order to design highly efficient CRS. However, only a few studies have been concerned with, on the one hand, the effect of the axial spacing, and, on the other hand, the design method \cite{Leesang}, particularly with rotors load distribution for a specified design point.

This paper focuses on two major parameters of ducted counter-rotating axial-flow fans in subsonic regime: the rotation rate ratio, $\theta$ and the relative axial spacing, $A$. In some cases, these systems are studied by using two identical rotors or the RR is not specifically designed to operate with the FR. In this study, the FR is designed as conventional rotor and the RR is designed on purpose to work with the FR at very small axial spacing. In this first design, The total work to perform by the CRS was arbitrarily set up approximately to two halves one half respectively for the FR and RR.
In \S~\ref{sec:design} the method that has been used to design the front and the rear rotors is firstly described. The experimental set-up is presented in \S~\ref{sec:setup}. Then the overall performances of the system in its default configuration and the effects of varying the rotation ratio and the relative axial spacing between the rotors are discussed in \S~\ref{sec:results}.

\section{\label{sec:design}Design of the rotors}
\subsection{General approach}
The design of the rotors is based on the use of the software MFT (Mixed Flow Turbomachinery), a 1D code developed by the DynFluid Laboratory \cite{Noguera93} ---based on the inverse method with simplified radial equilibrium--- to which an original method has been added specifically for the design of the RR of the counter-rotating system. 

From the specified total pressure rise, volume flow-rate and rotating speed, optimal values of the radii $R_{tip}$ and $R_{hub}$ are first proposed. In a second step, the tip and the hub radii as well as the radial distribution of the circumferential component of the velocity at the rotor outlet, $C_{u2}(r)$, could be changed by the user. The available vortex models are the free vortex ($C_{u2}(r)=\frac{K}{r}$), the constant vortex ($C_{u2}(r)=K^{'}$) and the forced vortex ($C_{u2}(r)=r K^{''}$).

The velocity triangles are then computed for $11$ radial sections, based on the Euler equation for perfect fluid with a rough estimate of the efficiency of $\eta_{est}=60\%$ and on the equation of simplified radial equilibrium (radial momentum conservation). The blades can then be defined by the local resolution of an inverse problem considering a 2D flow and searching for the best suited cascade to the proposed velocity triangles 
by the following parameters: $\gamma$ the stagger angle, computed from the incidence angle, $a$ giving the lower pressure variation on the suction surface of the blade using equations \ref{eq:gamma} and \ref{eq:a}. The solidity, $\sigma$ and the chord length, $c$ are thus computed at the hub and at the tip using equations \ref{eq:sigma} and \ref{eq:c} where $\cal D$ denotes the Lieblein's diffusion factor\cite{lieblein1953}. The intermediate chords are obtained by linearisation. Finally, the camber coefficients $C_{z\infty0}$ are computed using equation \ref{eq:coef_port}.
\begin{eqnarray}
	\gamma       & = & \beta_{1}-a \label{eq:gamma}\\
	a            & = & \frac{\Delta\beta+0.94}{q(\beta_{1})}+2.07 \label{eq:a}\\
	q(\beta_{1}) & = & 2.103-4.019 10^{-7} \beta_{1}^{3.382}\\
	\sigma^{-1}  & = & \left( {\cal D}-1+\frac{C_{2}}{C_{1}} \right) \times \left( \frac{2C_{1}}{|\Delta C_{u}|} \right) \label{eq:sigma}\\
	c            & = & \sigma \frac{2 \pi R}{Z} \label{eq:c}\\
	C_{z\infty0} & = & \frac{a+2.525}{p(\sigma)}-0.823 \label{eq:coef_port}\\
	p(\sigma)    & = & 15.535-12.467 e^{-0.4242 \sigma}
\end{eqnarray}
These empirical equations have been validated for NACA-65 cascades \cite{Noguera93}, for $0.5 \leq \sigma \leq 1.5$ and $0 \leq C_{z\infty0} \leq 2.7$.



\begin{figure}[t]
    \begin{center}
\includegraphics[clip,width=78mm]{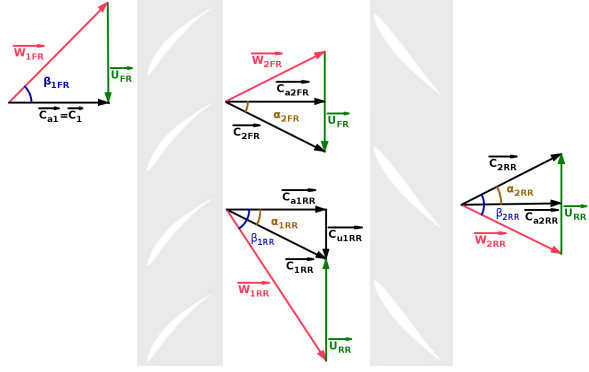}
    \end{center}
    \caption{\label{fig:TriangleVit}Velocity Triangles for the CRS. The fluid is flowing from left to right. 
    }     
\end{figure}

The behaviour of the designed machine resulting from the above method can then be analysed using a direct method in order to determine whether the design point is achieved and what are the characteristics of the machine at the neighbourhood of the design point.
The effects due to real fluid are taken partially into account with in-house loss models and the introduction of an axial-velocity distribution which considers the boundary layers at the hub and casing. Thus, the characteristics of the machine can be obtained in the vicinity of the design-point discharge.

Regarding the CRS, the geometrical dimensions, the number of blades of FR and of RR and their rotation rates are imposed. In particular, the number of blades of each rotor was chosen in order to prevent to have the same blade passing frequency or harmonics for both rotors in the lower frequencies range. The system that is presented here has moreover been designed to have a pure axial exit-flow.
An iterative procedure is then performed. The pressure rise of the FR is initially chosen and then designed and quickly analysed as explained. An estimate of the pressure rise that RR would made is then performed, based on this analysis. If the total pressure rise of the CRS is not met, the design pressure rise of FR is varied and the calculus are made again. In this method, losses and interactions in-between the two rotors are not taken into account. Any recirculation happening near the blade passage or near the blade hub or tip is not predicted by MFT as it is based on simplified radial equilibrium.

\subsection{\label{subsec:frdesign}Design of the Front Rotor}
 
\begin{table}[t]
\caption{\label{tab:speci}Design point of the counter-rotating system for air at $\rho=1.21~kg.m^{-3}$}.
\begin{center}
\begin{tabular}{p{1.5cm} p{1.5cm} p{1.5cm} p{1.5cm}}
& & & \\ 
	  \hline
	  & CRS & FR & RR  \\
	  \hline
	  \hline
	  $D$ (mm) & 380 & 380 & 380 \\
	  \hline
	  $R_{tip}$ (mm) & 187.5 & 187.5 & 187.5 \\
	  \hline
	  $R_{hub}$ (mm) & 55 & 55 & 55 \\
	  \hline
	  $Z$ & - & 11 & 7 \\
	  \hline
	  $\Delta p_t$ (Pa) & 420 & 260 & 160  \\
	  \hline
	  $N$ (RPM) & 1900 & 2000 & 1800 \\
	  \hline
	  $Q$ (m$^3$.h$^{-1}$) & 3600 & 3600 & - \\ 
	  \hline
	  $\Omega$ & 2.46 & 3.71 & - \\ 
	  \hline
	  Other constraints & Axial exit-flow & Constant vortex & -\\
	  \hline
	  \end{tabular}
\end{center}
\end{table}


The design point of the CRS is given in Tab.~\ref{tab:speci}. The system is designed to achieve a total-pressure rise $\Delta p_t=420$~Pa at flow-rate $Q=3600$~m$^3$.h$^{-1}$ for a rotation rate around $2000$~rpm ($\Omega \simeq 2.46$). The geometrical dimensions are fixed to fit in the set-up: $R_{tip}=187.5~mm$ and $\frac{R_{hub}}{R_{tip}}=0.293$. In this first and simple design, the front rotor (FR) has arbitrarily larger total pressure rise than the RR. The constant vortex model leads to a uniform tangential velocity distribution downstream FR for perfect fluid and thus has been used to simplify the design of the RR. Also, the rotors are not shrouded and the radial gap between the blade tip and the wall casing is of $2.5$~mm, \emph{i.e} $1.9\%$ of the blade height.

In the present case, the direct analysis predicts a mean absolute tangential velocity at the design flow-rate $C_{\theta 2FR} \simeq 9.6$~m.s$^{-1}$ with a radial distribution uniform within $\pm 5\%$ (constant vortex design). The Reynolds number based on the inlet relative velocity and the chord varies from $0.6 \times 10^5$ at the hub to $3 \times 10^5$ at mid-span and $7 \times 10^5$ at the tip.

\subsection{\label{subsec:rrdesign}Design of the Rear Rotor}
The method used for the design of the RR is to consider the velocity and the flow angle at the trailing edge of the FR blades. Therefore, FR was analysed with MFT to retrieve the axial and tangential velocities ($C_{a2FR}=C_{a1RR}$ and $C_{u2FR}=V_{u1RR}$ respectively) and the angle $\alpha_{2R1}$ in the absolute reference frame, at the exit and along the blade as shown in Fig.~\ref{fig:TriangleVit}. Therefore, the Euler work distribution along the blade does not match with any of the vortex models, previously mentioned.

\begin{figure}[hp]
    \begin{center}
    \includegraphics[clip,width=80mm]{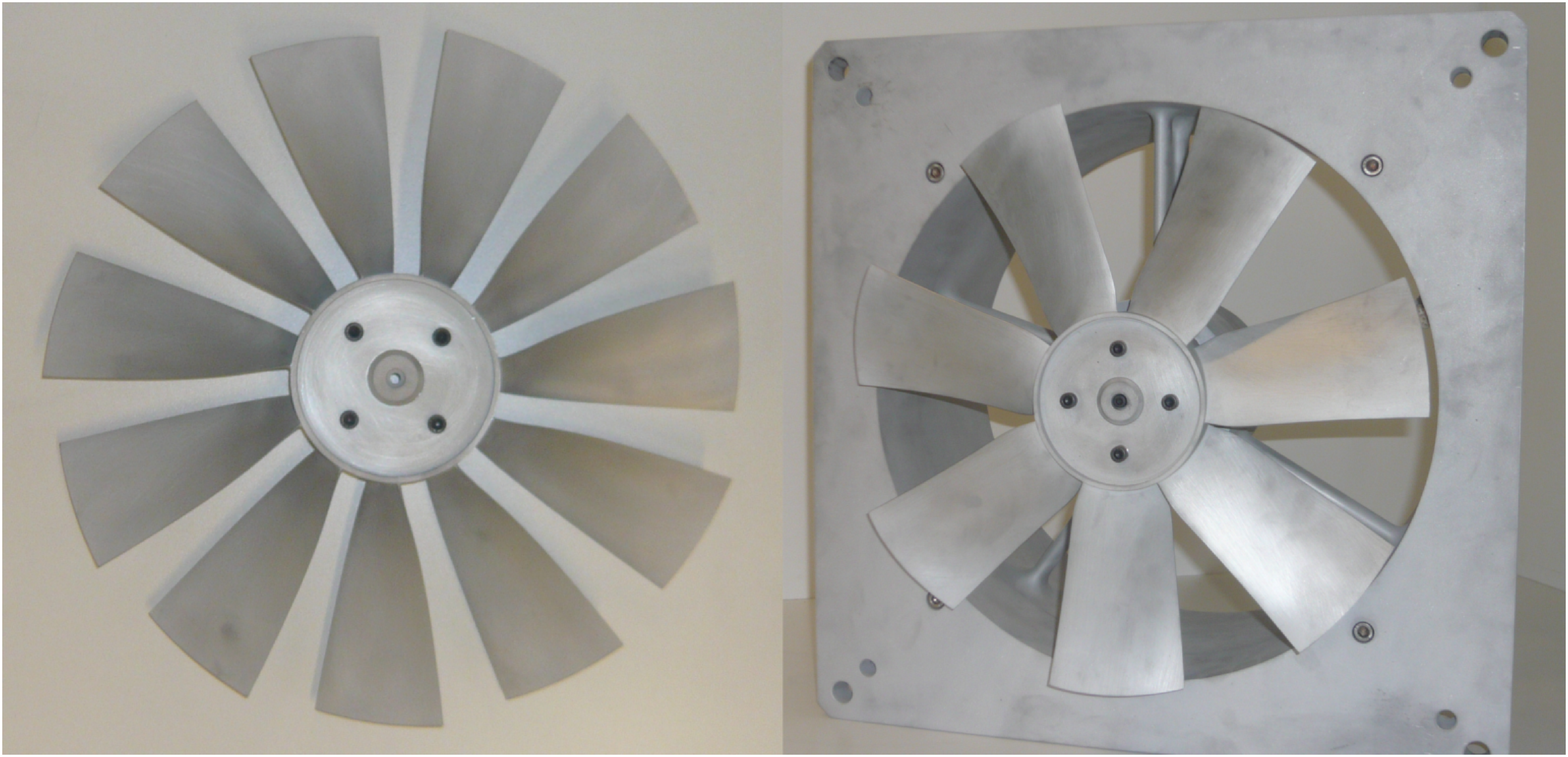}
    \includegraphics[clip,height=0.75\textheight]{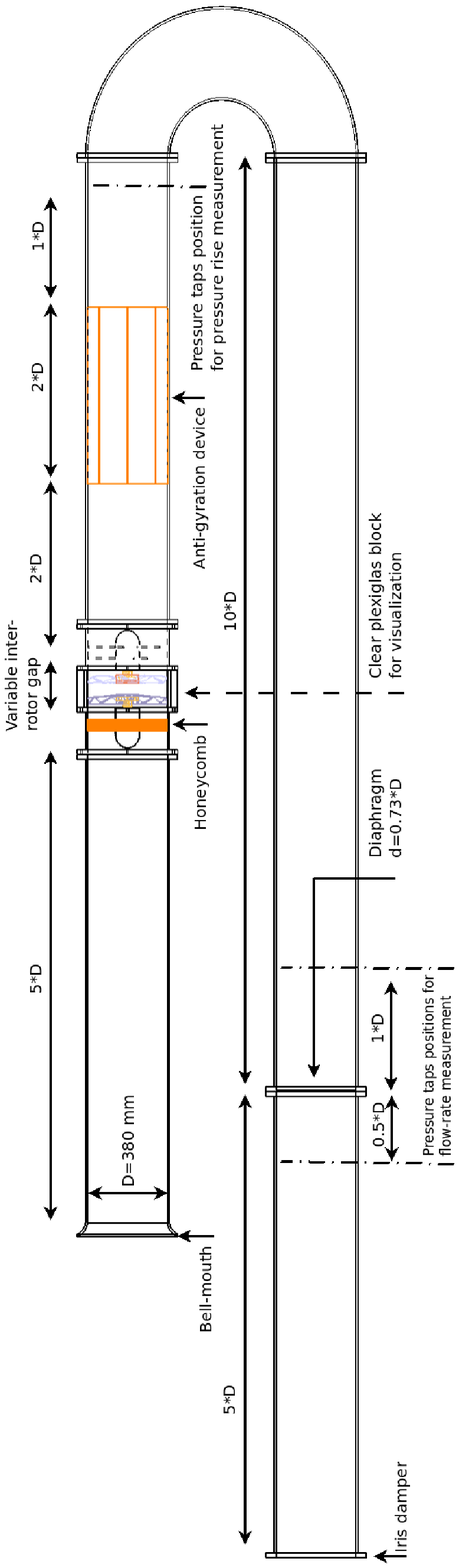}    
    \end{center}
    \caption{Top: picture of the front rotor (left) and Rear Rotor (right). Bottom: experimental facility for CRS, \it{AERO$^2$FANS}}
    \label{fig:HSN1_2} 
\end{figure}

Using the same radial  inverse design equations on 11 radial sections, the RR is drawn in such a way that the exit flow is purely axial, that is $\alpha_{2RR}(r)=0^o$. The second hypothesis is that the axial velocity profile is kept constant across RR, \emph{i.e.} $C_{a2RR}(r)=C0_{a1RR}(r)$. Under these assumptions, the total pressure rise of RR should be $\Delta p_{tRR} = \eta_{est} \, \rho \, U_{mRR} \, C_{u2FR} \simeq 0.6 \times 1.2 \times 22.9\times9.6 \simeq 160$~Pa where $\eta_{est}=0.6$ is an empirical estimated efficiency observed from previous experimental designs.
The blade cascade that lead to the desired velocity triangles is then designed with the previously described inverse method, adjusting the free parameters in such a way that the solidity lays in the range $0.5 \leq \sigma \leq 1.5$ and that the camber lays in the range $0 \leq C_{z\infty0} \leq 2.7$. After several iterations, the RR was drawn with $Z=7$, ${\cal D}_{hub}=0.61$ and ${\cal D}_{tip}=0.46$.
The geometrical characteristics of the rotor blades obtained with this method are summarized in Tab.~\ref{tab:geospeci}. Pictures of the Front and Rear rotors are given in Fig.~\ref{fig:HSN1_2}.

\begin{table}[t]
\caption{\label{tab:geospeci}Blade cascade parameters for the two rotors. Radius $R$ (mm). Chord length $c$ (mm). Cascade solidity $\sigma$. Stagger angle $\gamma$ ($^o$). Profile designation according to the nomenclature given in Ref.~\cite{Noguera93}: NACA65(xx)yy with (xx) representing the relative camber and yy standing for the relative thickness. Lieblein's diffusion factor $\cal D$}
\begin{tabular}{p{1.cm} p{0.7cm} p{0.4cm} p{0.4cm} p{0.2cm} p{2.35cm} p{0.35cm}}
 & & & & & & \\ 
\hline
		   & $R$ & $c$ & $\sigma$ & $\gamma$ & profile & $\cal D$\\
\hline
\hline
\multicolumn{7}{c}{}\\
\multicolumn{7}{c}{Front Rotor (blade thickness $4.5$~mm)}\\
\hline
Hub & $55$ & $40.3$ & $1.28$ & $23$ & NACA 65(26)11 & $0.62$ \\
\hline
Mid-span & $121.25$ & $58.0$ & $0.84$ & $57$ & NACA 65(12)07 & \\
\hline
Tip & $187.5$ & $75.7$ & $0.71$ & $69$ & NACA 65(07)06 & $0.44$ \\
\hline
\multicolumn{7}{c}{}\\
\multicolumn{7}{c}{Rear Rotor (blade thickness $6$~mm)}\\
\hline
Hub & $55$ & $58.8$ & $1.18$ & $73$ & NACA 65(03)10 & $0.61$ \\
\hline
Mid-span & $121.25$ & $72.9$ & $0.66$ & $65$ & NACA 65(05)08 & \\
\hline
Tip & $187.5$ & $87.1$ & $0.51$ & $75$ & NACA 65(04)07 & $0.46$ \\
\hline
\hline
\end{tabular}
\end{table}

\section{\label{sec:setup}Experimental set-up}
	\subsection{Test bench}
The counter-rotating system is studied in a ducted-flow test rig ---AERO$^2$FANS--- that has been built according to the ISO-5801 standards (installation of category B) \cite{iso5801,Ravelet2010}. The Figure~\ref{fig:HSN1_2} shows this test rig. Two brushless PANASONIC A4 motors drive each rotor separately and are hidden in a casing of diameter $0.33 D$ and of length $0.45 D$, with a warhead-shape end. The front and the rear motors are bound to the tube by two rod rows ($3$ and $5$ rods, the first row being at $0.1 D$ from the RR). For the front motor a honeycomb is placed between the two rows to reduce the rods effect on the incoming flow at the inlet of the FR and to homogenize the inlet flow. An iris damper ---originally used for air flow regulation in ducts--- is placed at the exit of the pipe to vary the test-bench hydraulic impedance and thereby to vary the operating point of the studied axial-flow fan. Finally, an axial blower can also be used at the exit of the pipe to widen the explored flow-rate.  

	\subsection{Measurements methodology}
The study focuses on the influence of the relative axial spacing $A=\frac{S}{c_{FRm}}$ where $S$ and $c_{FRm}$ are the axial spacing and the FR chord length at mid-span respectively, as well as on the influence of the rotation rate ratio $\theta=\frac{N_{RR}}{N_{FR}}$.
Six axial spacings, from $A=0.17$ to $A=3.1$ are investigated by means of blocks of different thickness. The reason of positioning RR the closest to FR and then increasing $A$ is to investigate on any possible potential effects due to the rotors proximity. Regarding the rotation rate ratio, each rotor is driven separately so all combinations are possible and the only limitation is the motor power. Unless specified the default axial spacing is $A=0.17$ and the default rotation rate ratio is $\theta=0.9$ (see Tab.~\ref{tab:speci}).

The static pressure rise of the rotor is obtained according to the ISO-5801 standards by measuring the pressure difference between the atmosphere and downstream the anti-gyration device.
The static pressure losses mainly induced by the honeycomb, the motors casings and the anti-gyration device have been measured using an auxiliary axial blower (with both rotors removed) and have been added to the static pressure rise measurements.

In the design of the CRS, it has been imposed a pure axial-flow at the exit of RR. In that case the static pressure rise of the CRS should be $\Delta p_s=\Delta p_t \; - \; 1/2 \, \rho \, \left( Q/(\pi D^2/4) \right)^2 \simeq 373$~Pa.
If the flow at the exit is not purely axial, then the total pressure rise will remain the same but the static pressure rise will be smaller and equal to: $\Delta p_s=\Delta p_t \; - \; \langle \, 1/2 \, \rho \, \vec{C_2(r)}^2 \, \rangle$.

The static efficiency is defined by equation \ref{eq:Efficiency}:
\begin{equation}
  \eta_s = \frac{\Delta P_s Q}{(T_{FR} \omega_{FR})+(T_{RR} \omega_{RR})}
  \label{eq:Efficiency}
\end{equation}
The torque $T$ was measured using the drivers provided with the motors. A calibration measurement has been performed with a conventional torque-meter. This calibration step shows that the torque supplied by the driver is very close to that given by the torque-meter (relative error of $0.5\%$).

Finally, for all performance measurements, the atmospheric pressure, the dry temperature and the wet temperature were measured and thus the density was computed for each measurement. It has been found that the variation in density, relatively to the design density, $\rho_{air}=1.21~kg.m^{-3}$ is between $0.5\%$ and $2.2\%$. Therefore, in order to present homogeneous results, the pressure rise is multiplied by the ratio of design density $\rho_{air}$ over the experimentally measured density, $\rho_{exp}$, \emph{i.e.} $\frac{\rho_{air}}{\rho_{exp}}$. 

\section{\label{sec:results}Results and discussion}
 
\subsection{\label{subsec:defaut}Overall performances of the reference system ($\{\theta=0.9 \, ; \, A=0.17\}$)}

\begin{figure}[t]
    \begin{center}
    \includegraphics[clip,width=0.99\columnwidth]{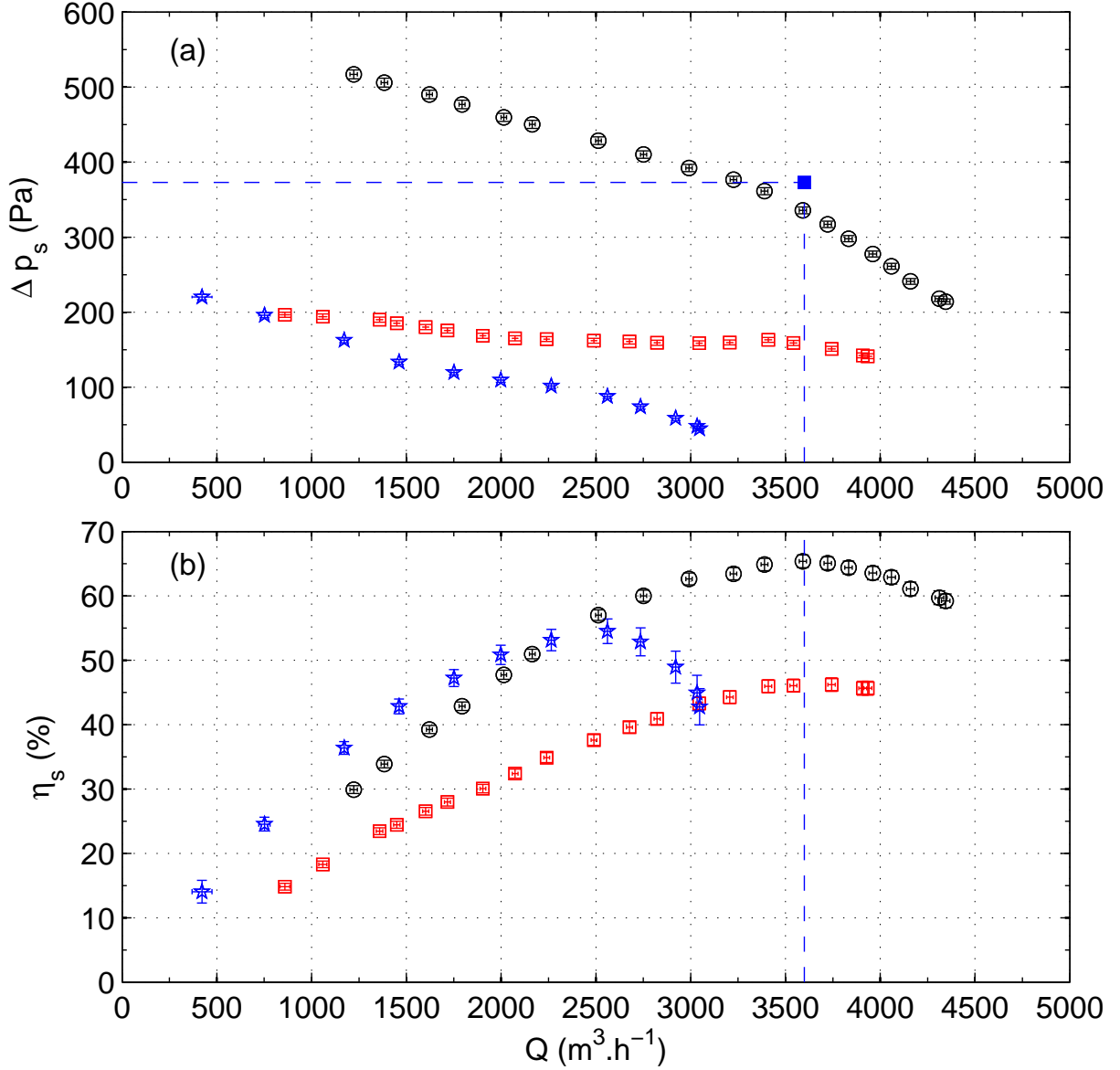}
    \end{center}
    \caption{\label{fig:caract_R1_R2_R12_nominal}Fans characteristics: (a) static pressure rise $\Delta p_s$ \emph{vs} flow rate $Q$; (b) static efficiency $\eta_s$ \emph{vs} flow rate $Q$. The axial spacing is $A=0.17$. $\Box$: FR rotating alone at $N_{FR}=2000$~rpm (RR has been removed), $\bigstar$: RR rotating alone at $N_{RR}=1800$~rpm (FR has been removed) and $\circ$: CRS at $N_{FR}=2000$~rpm and $\theta=0.9$. The $\blacksquare$ and the dashed lines stand for the design point of the CRS}
\end{figure}

The characteristics of the FR rotating alone (RR has been removed from its shaft in that case), of the RR rotating alone (FR has been removed) and of the counter-rotating system are shown in Fig.~\ref{fig:caract_R1_R2_R12_nominal}.

The nominal flow-rates of the three systems, \emph{i.e.} the flow-rates at maximum efficiency, are reported in Tab.~\ref{tab:rendMax} together with the corresponding static pressure rises and efficiencies. 

\begin{table}[t]
\caption{\label{tab:rendMax}Nominal points of FR rotating alone at $N_{FR}=2000$~rpm, RR rotating alone at $N_{RR}=1800$~rpm and CRS at $N_{FR}=2000$~rpm and $\theta=0.9$ (see also Fig.~\ref{fig:caract_R1_R2_R12_nominal})}
    \begin{center}
\begin{tabular}{p{2cm} p{1.3cm} p{1.3cm} p{1.3cm}}
& & &\\ 
\hline
	  & FR & RR & CRS\\
	  \hline
	  \hline
	  Max efficiency ($\%$) & 46.2$\pm1\%$ & 54.4$\pm1\%$  & 65.1$\pm1\%$ \\
	  \hline
	  Nominal $Q$ (m$^3$.h$^{-1}$) & 3636$\pm36$ & 2520$\pm36$ & 3600$\pm36$ \\
	  \hline
	  $\Delta p_s$ (Pa) & 157$\pm3$ & 88$\pm3$ & 335$\pm5$\\
	  \hline
	\end{tabular}
  \end{center}  
\end{table}
The FR rotating alone has a very flat curve ($\Box$ in Fig.~\ref{fig:caract_R1_R2_R12_nominal}). 
The nominal flow-rate of FR is slightly greater than the design point ---it is $3\%$ greater. The measured static pressure rise at the design point is $157\pm3$~Pa, with a relatively low static efficiency of $46.2\%$. This is not surprising with no shroud and a large radial gap. Moreover, this is consistent with the estimated static pressure rise by MFT, which is around $151$~Pa.

The RR rotating alone has a steeper curve ($\bigstar$ in Fig.~\ref{fig:caract_R1_R2_R12_nominal})
 and its nominal flow-rate $Q \simeq 2600$~m$^3$.h$^{-1}$ is lower than the design flow-rate of FR and CRS. This is consistent with the bigger stagger angle of the blades (see Tab.~\ref{tab:geospeci}) and can be explained by examining the velocity triangles in Fig.~\ref{fig:TriangleVit} and considering the case with the FR coupled to the RR: the incoming velocity $C_{1RR}$=$C_{2FR}$ has an axial component as well as a tangential component. Hence, the flow angle in the relative reference frame reads:
\begin{equation}
    tan(\beta_{1RR})=\frac{U_{RR}+C_{u 1RR}}{C_{a1RR}}
    \label{eq:tanbeta}
\end{equation}
Now the case without the FR is considered and it is assumed that the flow through the honeycomb is axial. Since the tangential component does not exist any more, $C_{u1RR}=0~m.s^{-1}$. MFT estimates $\langle U_{RR} \rangle \simeq 22.9$~m.s$^{-1}$, $ \langle C_{a1RR} \rangle \simeq 8.8$~m.s$^{-1}$ and $ \langle C_{u1RR} \rangle= \langle C_{u2FR \rangle} \simeq 9.6$~m.s$^{-1}$, which leads to $\langle tan(\beta_{1RR}) \rangle \simeq 3.69$ at the blade mid-span. Supposing that RR rotating alone reaches its maximum efficiency for $\langle tan(\beta_{1RR}) \rangle \simeq 3.69$, equation~\ref{eq:tanbeta} implies that $\langle C_{a1RR} \rangle =\frac{\langle U_{RR} \rangle }{tan(\langle \beta_{1RR} \rangle )} \simeq 6.2$~m.s$^{-1}$, \emph{i.e.} $Q \simeq 2540$~m$^3$.h$^{-1}$. This is exactly the nominal flow-rate of RR rotating alone (see Fig.~\ref{fig:caract_R1_R2_R12_nominal} and Tab.~\ref{tab:rendMax}). It is clear from the above analysis why the nominal flow-rate of RR is lower than the design flow-rate.

The characteristic curve of the CRS ($\circ$ in Fig.~\ref{fig:caract_R1_R2_R12_nominal}) is steeper than the characteristic curve of FR. It is roughly parallel to the RR curve. The nominal flow-rate of the CRS matches well with the design flow-rate, \emph{i.e.} $1$~m$^3$.s$^{-1}$. The static pressure rise at the nominal discharge ($\Delta p_{sCRS}=335$~Pa) is $10.2\%$ lower than the design point ($373$~Pa), which is not so bad in view of the rough approximations used to design the system. Please notice that the static pressure rise of the CRS is not equal to the addition of the static pressure rise of the FR with the pressure static rise of the RR, taken separately.

The CRS has a high static efficiency ($\eta_{sCRS}=65\%$) compared to a conventional axial-flow fan or to a rotor-stator stage with similar dimensions, working at such Reynolds numbers \cite{bakir2002,bakir2003}. The gain in efficiency with respect to the FR is $20$ points, whilst an order of magnitude of the maximum gain using a stator is typically $10$ points\cite{bakir2002,bakir2003}.

Awaiting for more accurate local measurements of the flow angle at the exit of the CRS, a simple test of flow visualization with threads affixed downstream of the CRS was performed. It has been observed that without the RR the flow is very disorganized. When the RR is operating, at the design configuration ($\theta=0.9$ and $N_{FR}=2000~rpm$), the flow is less turbulent, the threads are oriented with a small angle at the exit. This small angle seems, however to decrease when $\theta$ is increased between $1$ and $1.1$. This is consistent with the results in section \ref{subsec:influence} where it is found that the nominal operating point is observed for a value of $\theta$ higher than the design value.

The flow-rate range for which the static efficiency lays in the range $60\% \leq \eta_s \leq 65\%$ is: $2750 \lesssim Q \lesssim 4150$~m$^3$.h$^{-1}$, that is from $76\%$ of the nominal flow-rate up to $115\%$ of the nominal flow-rate. One open question is to what extent the global performances of the CRS are affected by the axial spacing and the speed ratio, and whether the efficient range could be extended by varying the speed ratio.

\subsection{\label{subsec:influence}Influence of the rotation ratio $\theta$}

\begin{figure}[t]
    \begin{center}
 	\includegraphics[clip,width=0.99\columnwidth]{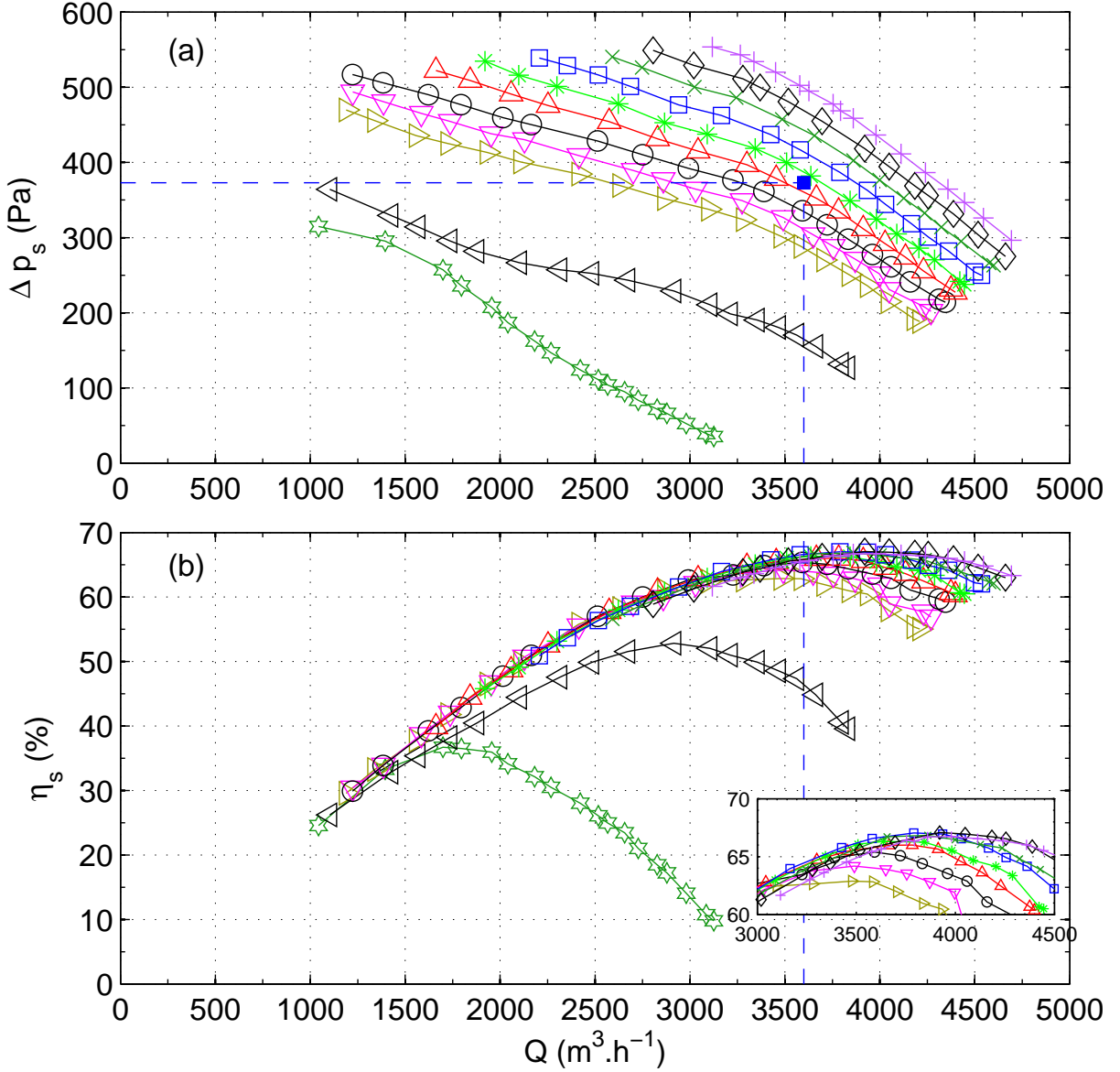}
    \end{center}
    \caption{\label{fig:caractRend_N1_2000} CRS characteristics at $N_{FR}=2000$~rpm, $A=0.17$ and $\theta \in [0 \, ; \,1.2]$~: (a) static pressure rise $\Delta p_s$ \emph{vs} flow rate $Q$; (b) static efficiency $\eta_s$ \emph{vs} flow rate $Q$. $\davidsstar$:
$\theta=0$, $\lhd$: $\theta=0.5$, $\rhd$: $\theta=0.8$, $\bigtriangledown$: $\theta=0.85$, $\diamond$: $\theta=0.9$, $\bigtriangleup$: $\theta=0.95$, $\ast$: $\theta=1$, $\Box$: $\theta=1.05$, $\times$: $\theta=1.1$, $\circ$: $\theta=1.15$ and $+$: $\theta=1.2$. The blue~$\blacksquare$ and the dashed lines stand for the design point of the CRS}
\end{figure}

\begin{figure}[t]
    \begin{center}
 	\includegraphics[clip,width=0.99\columnwidth]{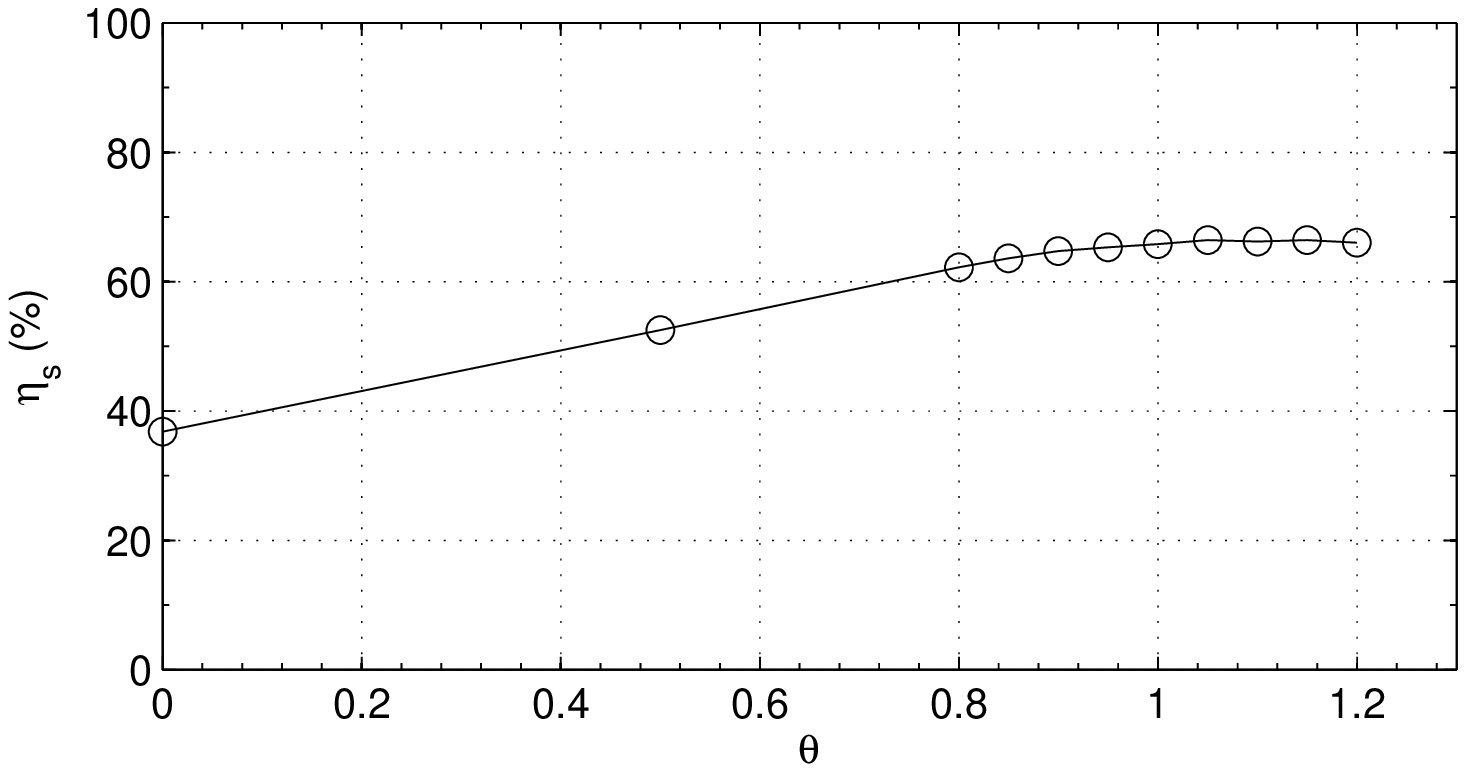}
    \end{center}
    \caption{\label{fig:etadetheta} Maximal static efficiency $\eta_s$ \emph{vs} $\theta$ for the CRS with $N_{FR}=2000$~rpm and $A=0.17$.}
\end{figure}

In this paragraph, the rotation rate of FR is kept constant at $N_{FR}=2000$~rpm, and the rotation rate of RR is varied from $0$ to $2400$~rpm. The corresponding $\theta$ are $\theta=\{0 \, ; \, 0.5 \, ; \, 0.8 \, ; \, 0.85 \, ; \, 0.9 \, ; \, 0.95 \, ; \, 1 \, ; \, 1.05 \, ; \, 1.1 \, ; \, 1.15 \, \& \, 1.2\}$. The axial spacing is $A=0.17$.

The overall performances of the CRS in these conditions are plotted in Fig.~\ref{fig:caractRend_N1_2000}. As expected, the more the rotation rate of RR increases, the more the static pressure rise of the CRS increases and the nominal flow-rate of the CRS increases. The maximal efficiency as a function of $\theta$ is plotted in Fig.~\ref{fig:etadetheta}.

For very low rotation rates of RR, \emph{i.e.} for $\theta=0$ ($\davidsstar$ in Fig.~\ref{fig:caractRend_N1_2000}) and $\theta=0.5$ ($\lhd$ in Fig.~\ref{fig:caractRend_N1_2000}), the system is very inefficient: in the first case when the RR is at rest the maximum efficiency hardly reaches $35\%$ which is below the maximal efficiencies of both FR and RR alone. The maximum flow-rate that can be reached is moreover very low in both cases compared to the discharge goal of $3600$~m$^3$.h$^{-1}$.

In the range $\theta \in [0.8 \, ; \,1.2]$, \emph{i.e.} $N_{RR} \in [1600 \, ; \,2400]$~rpm, the system is highly efficient. The maximum efficiency increases with $\theta$ to reach a maximum value of $66.5\%$ for $\theta=1.05$ and is then quasi-constant ($\eta_s=66.0\%$ for $\theta=1.20$). 

This is a very interesting feature of the counter-rotating system. One could imagine, simply by varying the RR rotation rate, to work at a constant pressure rise with an efficiency greater than $60\%$ for a large flow-rate range. For instance in the present case, the system could give a constant static pressure rise of $375$~Pa with $\eta_s \geq 60\%$ for $3000 \leq Q \leq 4250$~m$^3$.h$^{-1}$ with $N_{FR}=2000$~rpm, $A=0.17$ and $\theta \in [0.85 \, ; \,1.2]$. 

One could also imagine to work at a constant flow-rate with high static efficiency. For instance in the present case, the system could give a constant flow-rate of $3600$~m$^3$.h$^{-1}$ with $\eta_s \geq 60\%$ for $290 \leq \Delta p_s \leq 490$~Pa with $N_{FR}=2000$~rpm, $A=0.17$ and $\theta \in [0.8 \, ; \,1.2]$.

\subsection{Influence of the relative axial spacing $A$\\}
\begin{figure}[h]
    \begin{center}
    \includegraphics[clip,width=0.99\columnwidth]{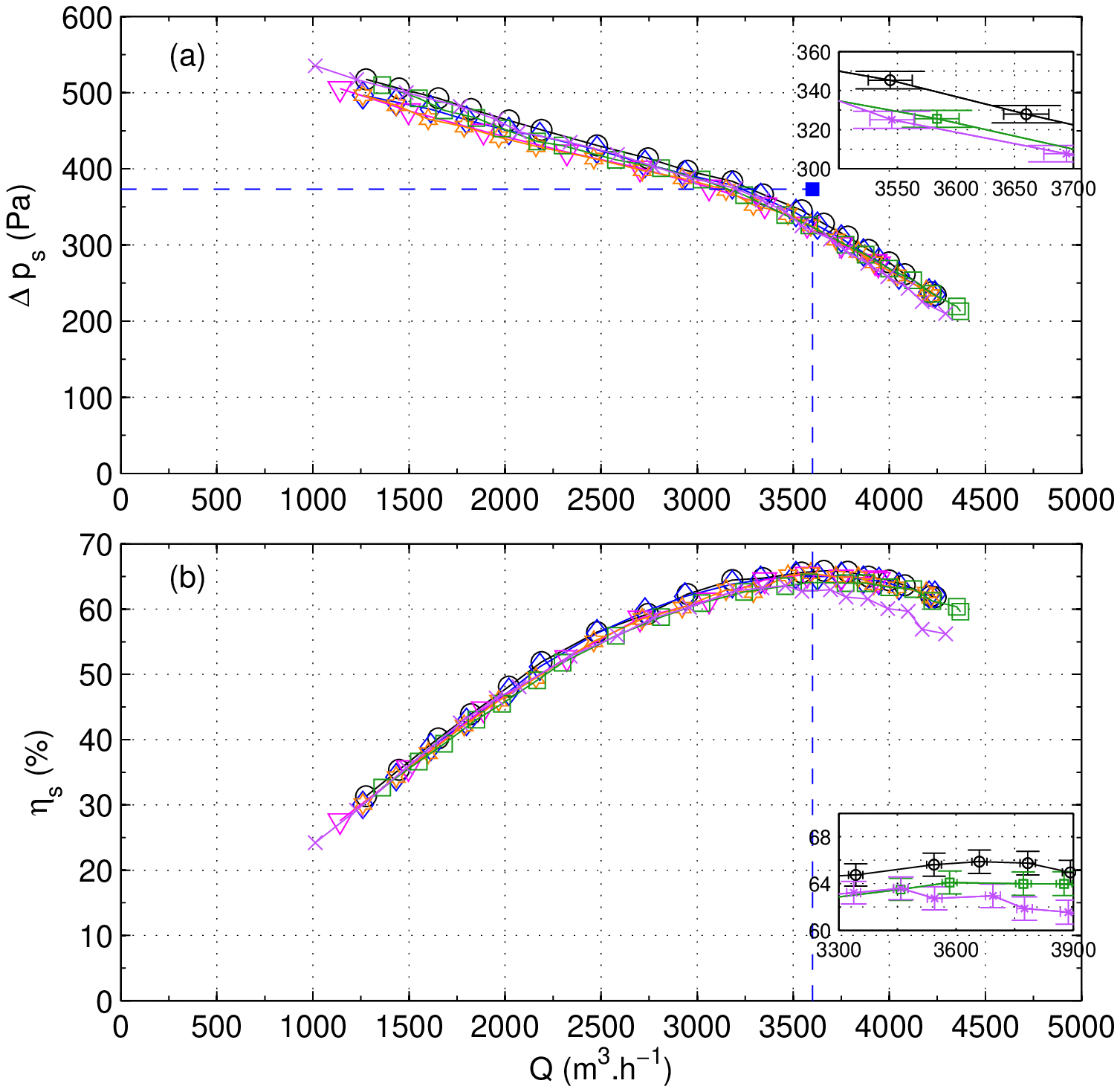}
    \end{center}
    \caption{\label{fig:caract_distance} CRS characteristics at various axial spacing: (a) static pressure rise $\Delta p_s$ \emph{vs} flow rate $Q$; (b) static efficiency $\eta_s$ \emph{vs} flow rate $Q$. The rotation ratio of FR is $N_{FR}=2000 rpm$ and $\theta=0.9$. $\circ$: $A=0.17$, $\diamond$: $A=.34$, $\triangledown$: $A=0.69$, $\davidsstar$: $A=0.86$, $\square$: $A=2.58$ and $\times$: $A=3.10$. The blue~$\blacksquare$ and the dashed lines stand for the design point of the CRS}
\end{figure}
Figure~\ref{fig:caract_distance} shows the characteristics curves at the design rotation rates, i.e., $N_{FR}=2000$~rpm and $\theta=0.9$. Regarding $A \in [0.17, \, 0.34, \, 0.69, \, 0.86]$, the overall performances do not change significantly and the variation is in the uncertainty range. The efficiency does not vary significantly either.

In other studies\cite{Sharma1988} it was reported that the axial spacing had a more significant influence on the overall performances. This was noticed as well in this study. For A=$2.58$ and A=$3.1$, the global performances are decreased by $\sim17$~Pa ($5\%$) comparing to the other spacings. However, even for A=$3.1$, the CRS still shows good performances with high efficiency compared to the conventional fan systems. 

\section{\label{sec:conclusion}Conclusion}
A counter-rotating  axial-flow fan has been designed according to an iterative method that is relatively fast. It is based on semi-empirical modelization that partly takes into account the losses, boundary layers at hub and casing, and the effects of \lq\lq low\rq\rq~ Reynolds numbers (below $2\times 10^5$).

The overall performances at the nominal design point are slightly lower than predicted, with a static pressure rise $10.2\%$ lower. The static efficiency is however remarkably high ($\eta_s \simeq 65\%$) and corresponds to a $20$ points gain in efficiency with respect to the FR maximal efficiency and to a $10$ points gain with respect to the RR. The overall measurements give first clues that allow to validate the design method.

The counter-rotating system has a very flexible use that allows to work at constant flow-rate on a wide range of static pressure rises or to work at constant pressure rise on a wide range of flow-rates, with static efficiency bigger than $60\%$, simply by varying the RR rotation rate. One could thus imagine an efficient closed-loop-controlled axial-flow fan. The overall performances moreover do not significantly vary with the axial spacing in the range $A \in [0.17 \, ; \, 0.86]$. However, for $A=2.58$ and $A=3.1$ the overall performances slightly decrease.


\end{document}